\documentclass{ws-procs9x6-cpt19}
\begin{document}

\newcommand{\refeq}[1]{(\ref{#1})}
\def\etal {{\it et al.}}

\title{The 2-Neutrino Exchange Potential with Mixing: \\ A Probe of Neutrino Physics and CP Violation}

\author{D.E.\ Krause$^{1,2}$ and Q.\ Le Thien$^1$}

\address{$^1$Physics Department, Wabash College, Crawfordsville, IN 47933, USA}

\address{$^2$Department of Physics and Astronomy, Purdue University, \\ West Lafayette, IN 47907, USA}

\begin{abstract}
The  2-neutrino exchange potential is a Standard-Model weak potential 
arising from the exchange of virtual neutrino--antineutrino pairs 
which must include all neutrino properties, 
including the number of flavors, 
their masses, 
fermionic nature (Dirac or Majorana), 
and CP violation.  
We describe a new approach for calculating 
the spin-independent 2-neutrino exchange potential, 
including the mixing of three neutrino mass states 
and CP violation.
\end{abstract}

\bodymatter

\phantom{}\vskip10pt\noindent
The neutrino sector of the Standard Model (SM)
holds  great potential for revealing new physics.  
Interestingly, 
the unsolved problems of neutrino physics 
[e.g., the masses of the three neutrino mass states,  
the neutrino's fermionic nature (Dirac or Majorana), 
number of flavors, 
existence of sterile neutrinos and CP violation] 
all impact the 2-neutrino exchange potential (2-NEP), 
the weak interaction force arising from 
the exchange of virtual neutrino--antineutrino pairs.  
The formulas for the single neutrino flavor 2-NEP 
were first derived by Feinberg and Sucher\cite{FS} 
and Fischbach\cite{Fischbach AoP} 
assuming massless and massive neutrinos, 
respectively.  
A number of other authors 
have also investigated the 2-NEP.\cite{Others,Grifols,Stadnik}  
Lusignoli and Petrarca\cite{LP} developed an integral formula 
for the 2-NEP with mixing of three neutrino flavors, 
but did not include all of the electroweak contributions. 
Here, 
we describe a new derivation of the 2-NEP with mixing 
that incorporates neutral-current (NC) 
and charged-current (CC) weak interactions 
and CP violation, 
and discuss the possibilities of using the 2-NEP 
as a probe of neutrino physics.\cite{LeThien}

In our approach for calculating the 2-NEP, 
we express the neutrino fields in the Schr\"{o}dinger picture 
and then use time-independent perturbation theory 
to calculate the second-order energy shift 
of the neutrino-field vacuum energy 
due to the presence of two stationary fermions.  
We ignore infinite self-energy corrections, 
which only depend on the positions of a single fermion.  
The spin-independent contribution, 
which depends on the separation distance $r$, 
is finite, 
and for the single-flavor case 
involving NC weak interactions 
is found to be given by\cite{LeThien}
\begin{equation}
V_{\nu,\bar{\nu}}(r) = \frac{G_{F}^{2}g_{V,1}^{f}g_{V,2}^{f}m_{\nu}^{3}}{4\pi^{3}r^{2}}K_{3}(2m_{\nu}r),
\label{1F 2-NEP}
\end{equation}
where $G_{F}$ is the Fermi constant, 
$g_{V}$ is the vector coupling constant, 
$m_{\nu}$ is the neutrino mass, 
and $K_{n}(x)$ is the modified Bessel function.

To incorporate mixing, 
we write the three flavor fields 
$\nu_{\alpha}(\vec{r})$, $(\alpha = e, \mu,\tau)$ 
as linear combinations of the three mass fields 
$ \nu_{a}(\vec{r})$, $(a = 1,2,3)$, 
i.e.,
$\nu_{\alpha}(\vec{r}) \equiv \sum_{a = 1}^{3}U_{\alpha a} \, \nu_{a}(\vec{r})$, 
where $U_{\alpha a}$ are components of the Pontecorvo--Maki--Nakagawa--Sakata (PMNS) matrix,
\begin{equation}
U_{\alpha a} = 
 \left(
\begin{array}{ccc}
c_{12}c_{13} & s_{12}c_{13} & s_{13}e^{-i\delta_{\rm CP}} \\
-s_{12}c_{23} - c_{12}s_{23}s_{13}e^{i\delta_{\rm CP}} &  c_{12}c_{23} - s_{12}s_{23}s_{13}e^{i\delta_{\rm CP}} & s_{23}c_{13} \\
s_{12}s_{23} - c_{12}c_{23}s_{13}e^{i\delta_{\rm CP}} & -c_{12}s_{23}-s_{12}c_{23}s_{13}e^{i\delta_{\rm CP}} & c_{23}c_{13}
\end{array}
\right).
\end{equation}
Here, $s_{ab} = \sin\theta_{ab}$, $c_{ab} = \cos\theta_{ab}$, and $\delta_{\rm CP}$ is the CP-violation phase. 
For the purpose of our calculation, 
we note that 
nucleons interact with the neutrino
only via NC interactions, 
while leptons also require the inclusion of CC interactions.  
We will therefore need to consider three cases for the interaction potentials: 
nucleon--nucleon, nucleon--lepton, and lepton--lepton.

The 2-NEP between nucleons is the simplest 
since the NC current interaction is independent of neutrino flavor.  
For nucleons \#1 and \#2, we merely sum Eq.~\refeq{1F 2-NEP}
over the three mass states, 
which gives\cite{LeThien}
\begin{equation}
V_{\rm N_{1},N_{2}}(r) = \frac{G_{F}^{2}g_{V,1}^{{\rm N}_{1}}g_{V,2}^{{\rm N}_{2}}}{4\pi^{3}r^{2}}\sum_{a = 1}^{3}m_{a}^{3}K_{3}(2m_{a}r),
\label{general nucleon V}
\end{equation}
where N = proton or neutron, 
$g_{V}^{{\rm N}} = \frac{1}{2} - 2\sin^{2}\theta_{W}$ for protons 
and $g_{V}^{{\rm N}} = -\frac{1}{2}$ for neutrons, 
where $\theta_{W}$ is the Weinberg angle.
The magnitude of this interaction is quite small. 
For two neutrons, 
the gravitational force is larger than 
the 2-NEP force when $r \gtrsim 1\,$nm.

For the case of a nucleon interacting with a lepton, 
one finds a result similar to the nucleon--nucleon potential 
except for a change of the lepton vector coupling, 
which depends on the PMNS matrix element 
corresponding to the lepton flavor,
\begin{equation}
V_{{\rm N}\alpha}(r) = \frac{G_{F}^{2}g_{V}^{\rm N}}{4\pi^{3}r^{2}}\sum_{a = 1}^{3}m_{a}^{3} \left(g_{V}^{\alpha} +|U_{\alpha a}|^{2} \right) K_{3}(2m_{a}r).
\label{general nucleon lepton V}
\end{equation}

The final case, 
the lepton--lepton 2-NEP, 
is the most interesting 
since the mixing has the greatest impact on 
the form of the potential 
and it involves both NC and CC interactions, 
but we were unable to obtain a closed-form expression 
if all the masses are nonzero.  
Instead, 
for two lepton flavors $\alpha$ and $\beta$, 
we found an expansion for the 2-NEP 
in powers of $(m_-^{ab}/m_+^{ab})$, 
where $m_{\pm}^{ab}  \equiv  m_a \pm m_b$:\cite{LeThien}
\begin{eqnarray}
V_{\alpha \beta}(r) & = &  \frac{G_{F}^{2}}{4\pi^{3}r^{2}}
\sum_{a = 1}^{3} \left[m_{a}^{3} \left(g_{V}^{\alpha} +|U_{\alpha a}|^{2} \right)\left(g_{V}^{\beta} +|U_{\beta a}|^{2} \right) K_{3}(2m_{a}r) \right]
\nonumber \\
&& \mbox{}+ V_{\rm {\alpha \beta, mix}}(r),
\label{E vac 2 lepton lepton massive}
\end{eqnarray}
where to order $ \left[ \left(\frac{m_-^{ab}}{m_+^{ab}} \right)^2 \right] $
\begin{equation}
\begin{split}
 V_{\rm {\alpha \beta, mix}}(r)  & \simeq    \frac{G_{F}^{2}}{4\pi^{3}r^{2}}
 \left[ \sum_{a > b}^{3}  \frac{ {\rm Re}\,\left( U_{\alpha a}^* U_{\alpha b}^{}  U_{\beta b}^* U_{\beta a}^{} \right)  }{4}
 \right.
 \\
& \hspace{-.35in} 
\left.\left\{
m_+^{ab} \left[\left(m_+^{ab}\right)^2 + \left(m_-^{ab}\right)^2 \right] K_3\left(\left.m_+^{ab}\right. r\right)
- \frac{4 \left(m_-^{ab}\right)^2 }{r} K_2\left(m_+^{ab}\, r\right)
\right\}
\vphantom{\sum_{a > b}^{3}  \frac{ {\rm Re}\,\left( U_{\alpha a}^* U_{\alpha b}^{}  U_{\beta b}^* U_{\beta a}^{} \right)  }{4}}\right].
\end{split}
\label{E vac 2 lepton lepton mixing massive}
\end{equation}
The lepton--lepton 2-NEP is particularly interesting 
because of its dependence on the CP-violating phase $\delta_{\rm CP}$.  Presently, 
there is growing evidence\cite{Abe CP} that 
$\delta_{\rm CP} \neq 0$.  
We show that 
one can write\cite{LeThien}
\begin{equation}
V_{\alpha\beta}(r) = V_{\alpha\beta}^{(0)}(r) + V_{\alpha\beta}^{({\rm CP})}(r)\, \sin^{2}\left(\frac{\delta_{\rm CP}}{2}\right),
\label{V CP}
\end{equation} 
where $V_{\alpha\beta}^{(0)}$ and $V_{\alpha\beta}^{({\rm CP})}(r)$ 
are complicated functions of $r$ 
independent of $\delta_{\rm CP}$, 
and $V_{ee}^{({\rm CP})}(r) = 0$ 
by the definition of $\delta_{\rm CP}$.  
Therefore, 
except for the interaction between two electrons, 
the 2-NEP potential will depend on the CP-violating phase 
due to the interference of the PMNS matrix elements.  

While the original study of 2-NEP 
arose mainly out of theoretical interests, 
our results raise the possibility of opening new avenues 
for experimental explorations of basic neutrino parameters.  
Because the 2-NEP involves the exchange of virtual neutrinos, 
all neutrino properties and energies must contribute.  
Besides being sensitive to the mixing angles, 
the 2-NEP depends directly on the actual neutrino masses, 
not the difference in mass squared 
as in neutrino-oscillation experiments.  
If we assume the neutrino mass spectrum 
lies in the range $1\,$meV $\lesssim m_{\nu} \lesssim 1\,$eV, 
the most promising experiments need to focus on 
the corresponding separations $1\,$nm $\lesssim r \lesssim 1\,\mu$m.    The dependence of the 2-NEP on CP violation is also novel, 
the only long-range SM force with this property.  
Muonium is the natural system to explore these effects 
since the 2-NEP CP violation requires interactions between leptons other than between just electrons. 
In addition, 
as pointed out by Fischbach \etal,  
the contribution of the 2-NEP to the nuclear binding energy 
provides an interesting test of the WEP 
as applied to neutrinos.\cite{Fischbach PRD} 
Of course, 
the experimental challenges in realizing these possibilities 
are significant, 
requiring measuring forces of less than gravitational strength 
on the nanometer scale 
or the Weak Equivalence Principle at the level of $\sim 10^{-17}$.  Recently, Stadnik\cite{Stadnik} 
examined the potential of using high-precision spectroscopy, 
although some of the assumptions made may be overly optimistic.\cite{Asaka}  

This work only touches on some of the interesting questions 
raised by the 2-NEP. 
We assumed neutrinos were Dirac fermions, 
examined only the spin-independent interaction, 
and assumed the simplest neutrino vacuum state.
Alternatives to these assumptions and others 
involving theories beyond the SM 
will certainly impact the 2-NEP, 
providing new directions for theoretical and experimental exploration.

\section*{Acknowledgments}
We thank Ephraim Fischbach for useful conversations 
and earlier papers on the 2-NEP, 
which provided significant motivation for our work.
We also thank Sheakha Aldaihan and Mike Snow 
for discussions on the derivation of potentials, 
which influenced our approach.


\begin{thebibliography}{xx}

\bibitem{FS} G.\ Feinberg and J.\ Sucher, Phys.\ Rev.\ {\bf 166}, 1638 (1968); G.\ Feinberg, J.\ Sucher, and C.-K.\ Au, Phys. Rep. {\bf 180}, 83 (1989).

\bibitem{Fischbach AoP} E.\ Fischbach, Ann.\ Phys. (NY) {\bf 247}, 213 (1996).

\bibitem{Others} J.B.\ Hartle, Phys. Rev. D {\bf 1}, 394 (1970); Phys. Rev. D {\bf 49}, 4951 (1994);   A.\ Segarra, arXiv:1606.05087; IOP Conf. Series: Journal of Physics: Conf. Series {\bf 888}, 012199 (2017); arXiv:1712.01049.

\bibitem{Grifols} J.A.\ Grifols, E.\ Mass\'{o}, and R.\ Toldr\'{a}, Phys. Lett. B {\bf 389}, 563 (1996).

\bibitem{Stadnik}  Y.V.\ Stadnik, Phys. Rev. Lett. {\bf 120}, 223202 (2018).

\bibitem{LP} M.\ Lusignoli and S.\ Petrarca, Eur. Phys.\ J.\ C {\bf 71}, 1568 (2011).

\bibitem{LeThien} Q.\ Le Thien and D.E.\ Krause, Phys.\ Rev.\ D (in press);  arXiv:1901.05345.

\bibitem{Abe CP}  K. Abe, \etal, Phys.\ Rev.\ Lett. {\bf 121}, 171802 (2018).

\bibitem{Fischbach PRD} E.\ Fischbach, D.E.\ Krause, C.\ Talmadge, and D.\ Tadi\'{c}, Phys. Rev. D {\bf 52}, 5417 (1995).

\bibitem{Asaka}  T.\ Asaka, M.\ Tanaka, K.\ Tsumura, and M.\ Yoshimua, arXiv:1810.05429.

\end{thebibliography}
\end{document}